\RequirePackage{fix-cm}
\documentclass[smallextended]{svjour3}       
\smartqed  
\usepackage[dvipdfmx]{graphicx}
\usepackage{color}
\usepackage{soul}
\soulregister\cite7
\soulregister\ref7
\soulregister\pageref7

\begin{document}

\title{Ground-state properties for bilayer Kitaev model:
dimer expansion study}%


\author{Akihisa Koga \and Hiroyuki Tomishige \and Joji Nasu}


\institute{A. Koga \and H. Tomishige \and J. Nasu \at
              Department of Physics, Tokyo Institute of Technology,
  Meguro, Tokyo 152-8551, Japan \\
}

\date{Received: date / Accepted: date}

\maketitle

\begin{abstract}
We study ground state properties in the bilayer Kitaev model
by means of the dimer expansion.
The existence of parity symmetries in the system
reduces the computational cost significantly.
This allows us to expand the ground state energy and
interlayer spin-spin correlation up to 30th order
in the interdimer Kitaev coupling.
The numerical calculations clarify that the dimer singlet state is indeed
realized in the wide parameter region.
\end{abstract}

\section{Introduction}
\label{intro}
Nonmagnetic states and the quantum phase transition
have attracted much interest in frustrated quantum spin systems.
One of the interesting examples is
the two-dimensional orthogonal-dimer system~\cite{Shastry,Miyahara}, where
nonmagnetic dimer and plaquette states compete with
the antiferromagnetically ordered state~\cite{PhysRevLett.84.4461}.
Recently, the pressure-induced phase transition
between the nonmagnetic states~\cite{P1,P2,P3} has been observed
in the candidate compound $\rm SrCu_2(BO_3)_2$~\cite{PhysRevLett.82.3168},
which stimulates theoretical investigations on
the competition between nonmagnetic
states~\cite{PhysRevLett.84.4461,Takushima}.
Another interesting nonmagnetic state is the quantum spin liquid state
in the Kitaev model~\cite{Kitaev2006}, where
spin degrees of freedom are decoupled into itinerant Majorana fermions and
$Z_2$ fluxes~\cite{Kitaev2006,Feng2007,Chen2007,Chen2008,koga2018}.
Two energy scales for distinct degrees of freedom
yield interesting finite temperature properties such as
double peak structure in the specific heat, spin dynamics,
and thermal Hall effect 
at low temperatures~\cite{Nasu2015,Yamaji2016,Yoshitake2016,Nasu2017,suzuki2018pre,yamaji2018pre,Nakauchi}.
In our previous paper~\cite{Tomishige}, we have considered two Kitaev models
connected by the Heisenberg exchange couplings,
which is one of simple models
to discuss the effect of the interlayer coupling.
Then, numerical calculations have suggested that
the interlayer coupling induces the first order quantum phase transition
to a nonmagnetic dimer state.
Here, we demonstrate the detailed analysis for the dimer expansion
to clarify how stable the dimer singlet state is against the quantum spin
liquid state.

In this work,
we study the bilayer Kitaev model by means of the dimer expansion method, 
which is an approach from the state composed of interlayer dimer singlets,
and discuss the stability of the dimer singlet state in the system.
First, we explain
the first order inhomogeneous differential method~\cite{Guttmann},
where physical quantities are deduced from
the series coefficients obtained from the dimer expansion.
Then, we clarify that the dimer singlet state is indeed realized
in the wide parameter region.

The paper is organized as follows.
In \S\ref{sec:model}, we introduce the model Hamiltonian
on the bilayer Kitaev model.
Then, we show the numerical results obtained
by the dimer expansion and study the stability of the dimer singlet state.
A summary is provided in the last section.

\section{Model and Results}
\label{sec:model}
\begin{figure}[htb]
\centering
\includegraphics[width=9cm]{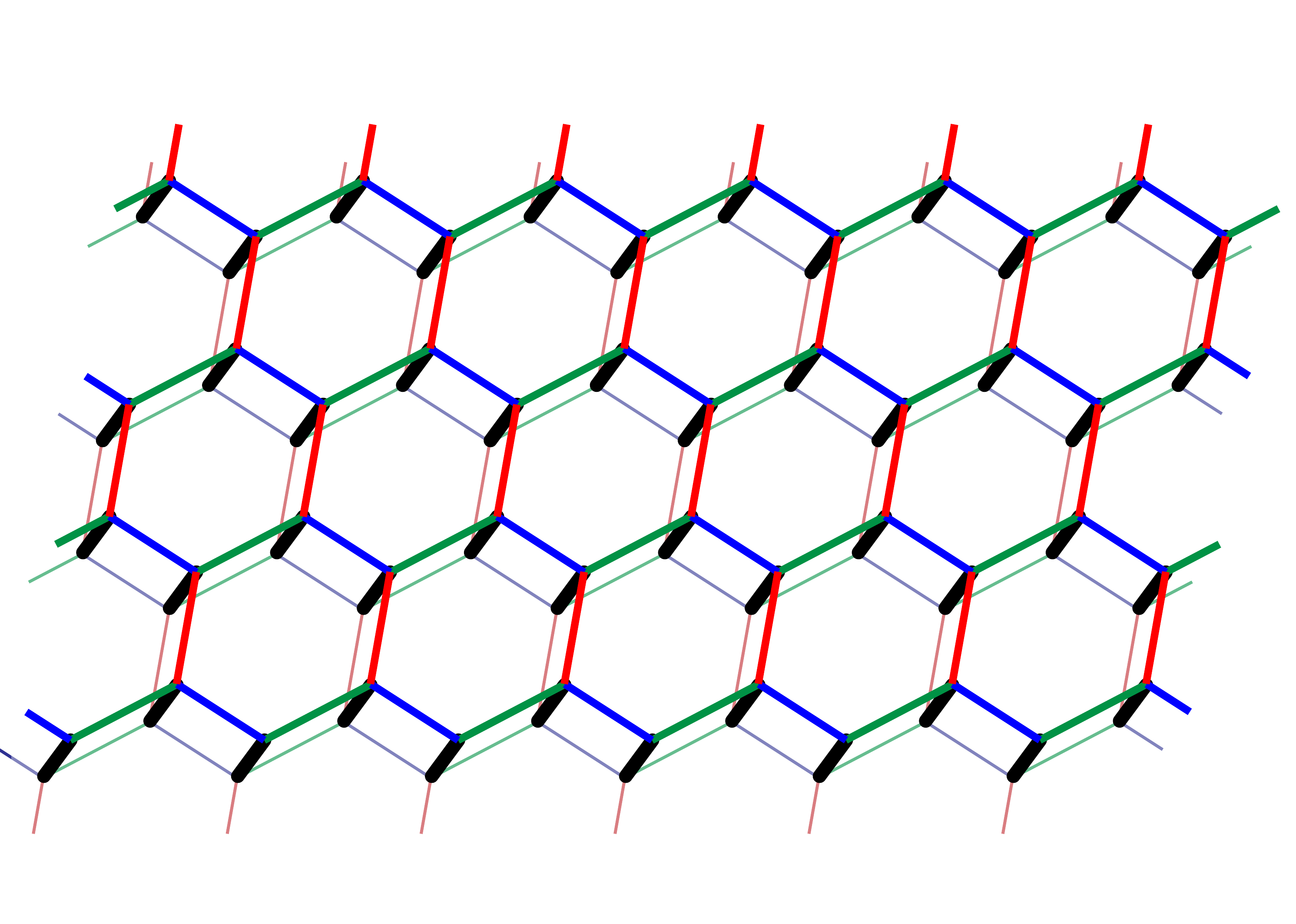}
\caption{
(Color online) Bilayer Kitaev model.
Green, blue, and red lines represent the Ising interaction
in the $x$, $y$, and $z$ direction,
and black lines the Heisenberg interaction. }
\label{fig:model}
\end{figure}

We consider the bilayer Kitaev model with the interlayer coupling,
which should be given by the following Hamiltonian as,
[see Fig.~\ref{fig:model}]:
\begin{eqnarray}
{\cal H}= -  J_K\sum_{\langle ij\rangle_\alpha,n}
S_{i,n}^\alpha S_{j,n}^\alpha
+J_H \sum_{i} {\bf S}_{i,1}\cdot{\bf S}_{i,2},\label{eq:1}
\end{eqnarray}
where $S_{i,n}^\alpha\; (\alpha=x,y,z)$ is the $S=1/2$ operator
at site $i$ of the $n(=1,2)$th layer,
$J_K (>0)$ is the ferromagnetic Kitaev coupling in each layer,
and $J_H (>0)$ is the antiferromagnetic Heisenberg coupling between two layers.
In each layer, the anisotropy of the Ising-type interactions
depend on the nearest neighbor bonds in the honeycomb lattice
(see Fig.~\ref{fig:model}).
When $J_H=0$, the system is reduced to two single-layer Kitaev models,
where the quantum spin liquid ground state is realized
with gapless excitations~\cite{Kitaev2006}.
On the other hand, in the case $J_K=0$,
the ground state is the direct product of
interlayer dimer singlets with the spin gap.
These two states have a difference in character, which should lead to
the quantum phase transition between the two limits.

To clarify the stability of the dimer singlet state,
we employ the dimer expansion technique~\cite{SE1,SE2,SE3}.
Since this method combines the conventional perturbation theory with
the cluster expansion, it has an advantage to deal with
the frustrated spin system in higher dimensions, where
the reliable results are hard to be obtained
by quantum Monte Carlo simulations.
In fact, using the dimer expansion, 
quantum phase transitions have been discussed
in the frustrated spin systems such as
the $J_1-J_2$~\cite{PhysRevB.63.104420},
orthogonal-dimer~\cite{PhysRevLett.84.4461,Takushima,Koga2000b},
and Kitaev-Heisenberg models~\cite{PhysRevB.96.144414}.

\begin{center}
\begin{table}
\caption{
Series coefficients for the dimer expansion of the ground state energy
$E_g/N=\sum a_i (J_K/J_H)^i$
and interlayer spin-spin correlation
$\langle {\bf S}_1\cdot{\bf S}_2\rangle =\sum b_i (J_K/J_H)^i$
in the bilayer Kitaev model.}
\begin{tabular}{c r r|c r r}
\hline\noalign{\smallskip}
$i$ & $a_i$ & $b_i$ &
$i$ & $a_i$ & $b_i$\\
\noalign{\smallskip}\hline\noalign{\smallskip}
0 & -0.75 & -0.75 &16 & 0.0725060850909 & -1.08759127636\\
2 & -0.1875 & 0.1875 & 18 & -0.0960012329502 & 1.63202096015\\
4 & 0.05859375 & -0.17578125 & 20 & 0.13044349648 & -2.47842643312\\
6 & -0.041015625 & 0.205078125 & 22 & -0.180983589717 & 3.80065538406\\
8 & 0.0376968383789 & -0.263877868652 & 24 & 0.255464209649 & -5.87567682193\\
10 & -0.0399161577225 & 0.359245419502 & 26 & -0.365835854119 & 9.14589635297\\
12 & 0.0461324302273 & -0.5074567325 & 28 & 0.530360009428 & -14.3197202546\\
14 & -0.0565949525058 & 0.735734382576 & 30 & -0.777038313098 & 22.5341110798\\
\noalign{\smallskip}\hline
\end{tabular}
\label{tab:II}
\end{table}
\end{center}
First, we divide the original Hamiltonian given by Eq.~(\ref{eq:1}) 
into two parts.
Since we start with the singlet state with strong $J_H$,
the second term of Eq.~(\ref{eq:1}) is considered as 
the unperturbed Hamiltonian ${\cal H}_0$, which is an assembly of interlayer
singlet-dimer formed by the coupling $J_H$~\cite{Hida1992,Hida}.
In this bilayer Kitaev model,
we use the following local basis sets given as
\begin{eqnarray}
 |s\rangle =\frac{1}{\sqrt{2}}\left(|\uparrow_1\downarrow_2\rangle-|\downarrow_1\uparrow_2\rangle\right),&\;\;&
 |x\rangle =-\frac{1}{\sqrt{2}}\left(|\uparrow_1\uparrow_2\rangle-|\downarrow_1\downarrow_2\rangle\right),\\
 |y\rangle =\frac{i}{\sqrt{2}}\left(|\uparrow_1\uparrow_2\rangle+|\downarrow_1\downarrow_2\rangle\right),&\;\;&
 |z\rangle=\frac{1}{\sqrt{2}}\left(|\uparrow_1\downarrow_2\rangle+|\downarrow_1\uparrow_2\rangle\right),
\end{eqnarray}
where $|s\rangle$ is the singlet state and
$|\alpha\rangle\;(\alpha=x,y,z)$ is the triplet state.
Then, the corresponding eigenenergies for the local Hamiltonian
$J_H{\bf S}_1\cdot{\bf S}_2$ are
$E=-3J_H/4$ and $J_H/4$, respectively.
The interactions among independent dimers are taken into account
by series expansions in the perturbed Hamiltonian
${\cal H}_1(={\cal H}-{\cal H}_0)$.
We wish to note that there exist global parity symmetries
in the number of singlet and triplet states
in this basis set~\cite{Tomishige}.
Therefore, when a certain cluster
is concerned in the framework of the cluster expansion,
the net states are restricted in the subspace
with even number of triplet states for each component $\alpha$,
which significantly reduce computation costs.

By performing the perturbation expansion for 787,894 graphs,
we obtain the series coefficients for
the ground state energy $E_g=\langle {\cal H} \rangle$ and
spin-spin correlations $C_s=\langle {\bf S}_1\cdot {\bf S}_2\rangle$
up to the 30th order in $J_K/J_H$.
The results are explicitly shown in Table~\ref{tab:II}.
We find that the series coefficients appear only in the even orders
in $J_K/J_H$ and are alternating.
Therefore, the extrapolations are necessary
to deduce physical quantities in the large $J_K/J_H$ region.
Pad\'e approximation is one of the powerful methods,
where the function is approximated
by the fractional of polynomials~\cite{Guttmann}.
This method can access the intermediate region with $J_K/J_H\sim 1$, but
we could not access the vicinity of the quantum phase transition point
$(J_K/J_H \sim 17)$~\cite{Tomishige}.
In this study, we make use of the extended method:
the first-order inhomogeneous differential method~\cite{Guttmann}.
In the method, the function $y(x)$ is numerically evaluated by
the following differential equation as
\begin{eqnarray}
x P_{N_1}(x) y'(x)+Q_{N_0}(x)y(x)=R_L(x),
\end{eqnarray}
where $P_n(x), Q_n(x)$, and $R_n(x)$ are
the $n$th order polynomials of $x$.
Since our obtained series has finite coefficients in even orders,
we apply the method to $y(x)=q(x^2)$, where $x=J_K/J_H$ and $q=E_g, C_S$.
\begin{figure}[htb]
\centering
\includegraphics[width=9cm]{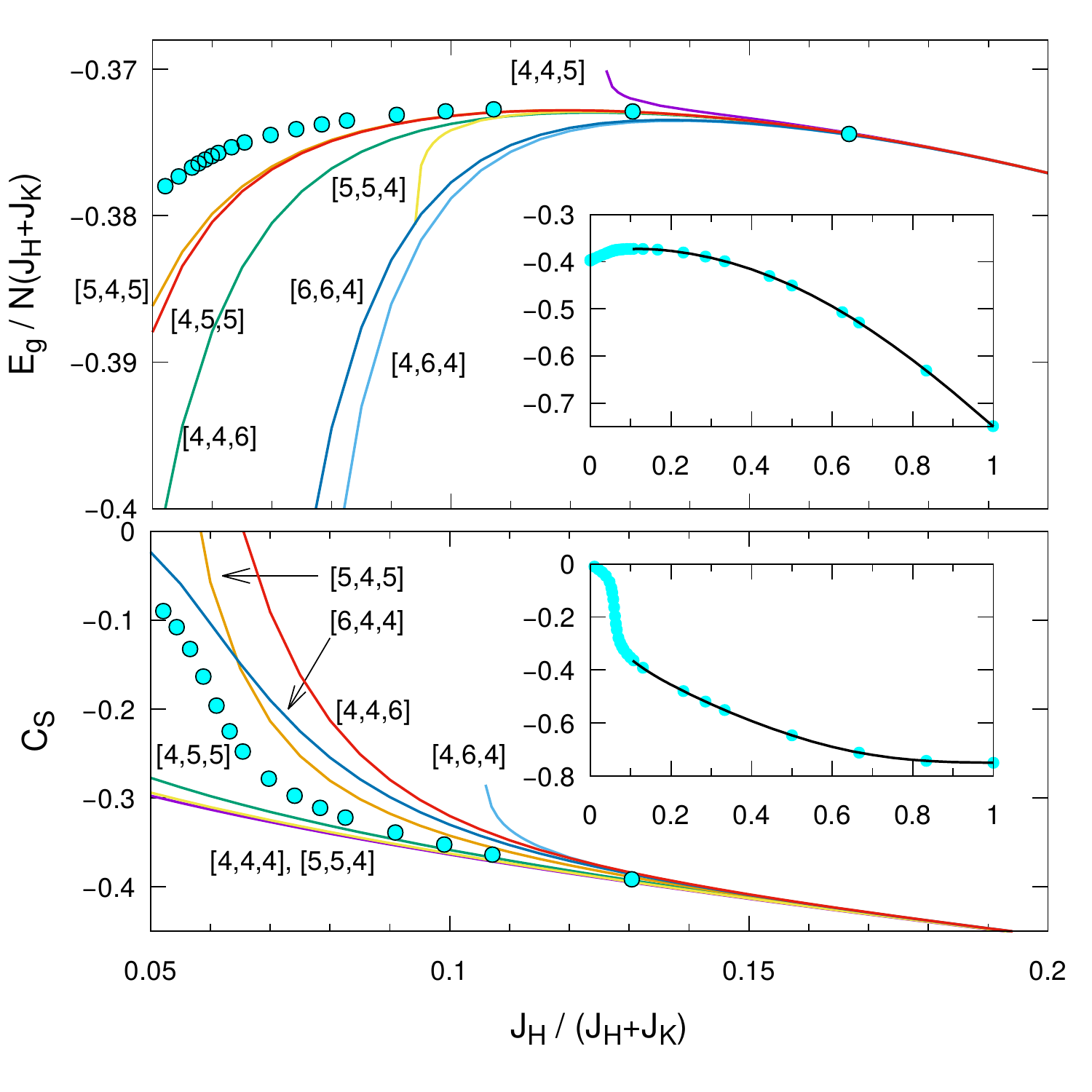}
\caption{
  (Color online) Ground state energy and spin correlations
  in the bilayer Kitaev model.
  Symbols are the results obtained by the exact diagonalization
  with $N=24$ sites~\cite{Tomishige}.
  Several lines with $[N_0,N_1,L]$ are obtained by
  the first-order inhomogeneous differential methods.
}
\label{fig:ene}
\end{figure}
Figure~\ref{fig:ene} shows the ground state energy and spin correlations
deduced by the first-order inhomogeneous differential method,
which is specified by $[N_0,N_1,L]$.
The extrapolated values show good agreement with
the results obtained by the exact diagonalization with $N=24$ sites
not only in the region of $J_K\sim 0$ but also $J_H/(J_H+J_K)>0.15$,
as shown in Fig.~\ref{fig:ene}.
Therefore, we can say that the dimer singlet state is indeed realized
in the region.
On the other hand, when $J_H/(J_H+J_K)\rightarrow 0.1$,
we find pathological singularities in some curves,
which sometimes occur in Pad\'e and related methods.
Although this makes it hard to deduce the physical quantities quantitatively
in the region $J_H/(J_H+J_K) < 0.15$,
one can expect that both quantities are smoothly changed.
This is in contrast to the ED results for spin-spin correlations,
where a rapid change appears around $J_H/(J_H+J_K)\sim 0.05$.
This implies that
the dimer expansion does not describe the ground state with
$J_H/(J_H+J_K)< 0.05$, which suggests the existence of
the first-order quantum phase transition between the dimer
and quantum spin liquid states.

\section{Summary}\label{sec:summary}
We have investigated ground state properties in the bilayer Kitaev model
by means of the dimer expansion.
By deducing the ground state energy and
spin correlation between layers,
we have confirmed that the dimer singlet state is indeed realized 
in the wide parameter region.
The detail of the dimer expansion has been addressed.

\begin{acknowledgements}
This work is supported by Grant-in-Aid for Scientific Research from
JSPS, KAKENHI Grant Nos. JP17K05536, JP18K04678 (A.K.) and
JP16K17747, JP16H02206, JP16H00987 (J.N.).
Parts of the numerical calculations were performed
in the supercomputing systems in ISSP, the University of Tokyo.
\end{acknowledgements}

\bibliographystyle{spphys}       
\bibliography{./refs}   

\end{document}